\documentclass{article}

\usepackage{arxiv}

\usepackage{graphicx}

\usepackage{amsmath,amsfonts,amssymb,amsthm, bm}
\usepackage{nccmath}
\usepackage{geometry}
\usepackage{algorithmicx,algpseudocode}
\usepackage{lineno}
\usepackage{textcomp}
\usepackage{xcolor}
\usepackage{natbib}
\usepackage{mathtools}
\setcitestyle{numbers,square}
\usepackage{subcaption}
\usepackage{braket}
\usepackage[ruled,vlined]{algorithm2e}

\usepackage[utf8]{inputenc} 
\usepackage[T1]{fontenc}    
\usepackage{hyperref}       
\usepackage{url}            
\usepackage{booktabs}       
\usepackage{nicefrac}       
\usepackage{microtype}      
\usepackage{lipsum}
\newcommand{\grad}{$^{\circ}$}
\newcommand*{\field}[1]{\mathbb{#1}}%

\title{A novel algorithm for confidence sub-contour box estimation: an alternative to traditional confidence intervals}

\author{
  Daniel Rojas-Diaz.\thanks{Corresponding author} \\
  Department of biological science\\
  school of basic science\\
  Universidad Eafit\\
  Medellin, Colombia \\
  \texttt{drojasd@eafit.edu.co} \\
   \And
    Alexandra Catano-Lopez.\\
  Department of mathematical science\\
  school of basic science\\
  Universidad Eafit\\
  Medellin, Colombia \\
  \texttt{acatano@eafit.edu.co} \\
  \And
    Carlos M. Velez-Sanchez\\
  Department of mathematical science\\
  school of basic science\\
  Universidad Eafit\\
  Medellin, Colombia \\
  \texttt{cmvelez@eafit.edu.co} \\
  }

\begin{document}
\maketitle

\begin{abstract}

The factor estimation process is a really challenging task for non-linear models. Even whether researchers manage to successfully estimate model factors, they still must estimate their confidence intervals, which could require a high computational cost to turn them into informative measures. Some methods in the literature attempt to estimate regions within the estimation search space where factors may jointly exist and fit the real data (confidence contours), however, its estimation process raises several issues as the number of factors increases. Hence, in this paper, we focus on the estimation of a subregion within the confidence contour that we called as Confidence Sub-contour Box (CSB). We proposed two main algorithms for CSB estimation, as well as its interpretation and validation. Given the way we estimated CSB, we expected and validated some useful properties of this new kind of confidence interval: a user-defined uncertainty level, asymmetrical intervals, sensitivity assessment related to the interval length for each factor, and the identification of true-influential factors.

\end{abstract}


\section{Introduction}
The development of mathematical models that describe real-life situations plays an important role in basic and applied research of disciplines as physics, biology, chemistry, engineering, and epidemiology \cite{Barrios2013}\cite{Canto2017}. Those models usually depend on several factors (parameters and initial conditions) whose values are unknown or only known imprecisely \cite{Voss2004}. Hence, the parameter estimation (factor estimation in this paper), as the assignment of values to the factors in a model to optimize the agreement between the model and real data, and the calculus of its confidence intervals (CI) are essential when modeling dynamical systems to study relations among variables \cite{Canto2017}\cite{Geris2016}. CI estimation is important for model validation, risk assessment, uncertainty analysis, and sensitivity analysis \cite{Saltelli2008}. However, for non-linear systems, CI achievement is a challenge because there may be several sets of factors that meet optimization criteria \cite{Sandia2005}\cite{Draper1998}. When the CI group several sub-intervals where the factors meet optimization criteria, they also enclose potentially wrong regions, i.e., an overestimation of ranges for the factors that might affect posterior analysis \cite{Sandia2005}. Thus, we will propose of a new kind of CI and its achievement for non-linear models focusing in a previous promissory result from factor estimation, even under uncertainty in real data or identifiability issues.

When proposing a model for factor estimation, the researchers must have some \textit{a priori} knowledge about the values that each factor must take \cite{Geris2016}; thus, it is likely possible to declare an interval for each parameter. On the other hand, we can represent any set of factors as a coordinate in a $n-$dimensional space (space of factors), where each axis represents one of the $n$ model factors. Then, through defining factor intervals, we are also creating a $n-$orthotope that englobes all feasible sets of factors (search box). Additionally, in real life, there is uncertainty in data; therefore, when we perform parameter estimation, we are not characterizing which points but regions of the search box are consistent with real data and prior knowledge, i.e., contours of the search box that encloses real data (confidence contour) \cite{Geris2016}.

For factor estimation, it is common to apply optimization methods that minimize a least-squares expression (coincidence between model output and real data) \cite{Sandia2005}. Thus, it is possible to map each point from search space into least squares surface, and then, characterize this surface through contours \cite{Draper1998}. For linear models, the surface contours would be ellipsoidal and would have a single global minimum height; while for non-linear models, the contours are not ellipsoidal but tend to be irregular, perhaps with several local minima and even more than one global minimum \cite{Draper1998}. After factor estimation task, it is highly recommended to estimate the CI to test and generate a hypothesis that helps to understand the processes underlying the real data \cite{Babtie2014}. The usual way to CI achievement is to perform a large number of parameter estimation tasks, and then, to calculate the probability distribution of each factor \cite{Draper1998}. However, as discussed previously, every interval defined separately for each factor defines a $n-$orthotope, while confidence contour has an ellipsoidal or irregular form; thereby, the CI may lead to large errors if one is trying to infer a region where the factors may jointly exist \cite{Sandia2005}.

There are some methods to estimate confidence contours instead of CI \cite{Seber2003}. Looking at the confidence contours will give a better picture of the optimal values for the factors, and their correlation \cite{Seber2003}. Unfortunately, the complexity in the calculus of confidence contours increases as the number of parameters rises \cite{Sandia2005}. Then, we do not focus on estimate confidence contours; since with large models, it is not worth, or not even possible to estimate them. On the other hand, multiple global or local minima may intricate the estimation of both of CI or confidence contour because of its relation with model identifiability \cite{Miao2011}. Thus, there is a problem with gradient search algorithms for factor estimation: since each initial point in factor estimation task can lead to convergence in a different local minimum, then resultant confidence intervals may enclose two different regions that fit with real data \cite{Sandia2005}. Those overestimated confidence intervals can be non-informative and lead to wrong conclusions, especially for uncertainty and sensitivity analysis \cite{Saltelli2008}.

In practice, factor and CI estimation for non-linear models is a difficult task due to the irregular form of confidence contour and the multiple local or global minima of its least-squares surface. Hence, in the present paper, we propose a method to calculate a different type of CI for a researcher-selected minimum. We also guarantee that the resultant $n-$orthotope is a subregion of the confidence contour for the minimum, therefore we call it the Confidence Sub-contour-Box (CSB); also, we managed to establish a direct relation between sensitivity concepts and CSB to achieve a unique box with some useful properties for the general purposes of model validation, risk assessment, and uncertainty analysis. \\

The paper is distributed as follows: first, we present a background for factors and confidence intervals; where we explain the basis of dissimilarity quantification, parameter estimation, and confidence intervals. Then, we present a section of the construction of a promissory search box, in which we define a threshold criterion to select a sub-contour box. In Section uncertainty-based confidence interval shrink, we present the algorithm that estimates a confidence sub-contour box beside its explanation. Finally, we present an application case for the algorithm using the novel GSUA-CSB toolbox and a dengue spread model.

\section{Background for estimation of factors and confidence intervals}\label{sec:background}
   In this section, we want to briefly introduce some approaches for dissimilarity quantification among model output vectors, its relevance for parameter estimation, and its relation with sensitivity analysis. This is important because the method we will define has its foundation in dissimilarity among model outputs regarding a previously-known nominal output. Further, we hypothesize that said dissimilarity must state a direct correlation with less feasible values for sensitive parameters in the model. \\

\begin{table}[htbp!]
   \centering
   \caption{Notations used in the text\label{Tab:definition}}
    \begin{tabular}{cl}
    \hline
    \textbf{Symbol} & \multicolumn{1}{c}{\textbf{Description}} \\ \hline
    $|\cdot|$ & Cardinality of argument $(\cdot)$ \\
    $N$ & Sample size \\
    $k$ & Number of factors \\
    $X$ & Vector of input factors $(1\times k)$. Also, it could represent a matrix of $(N\times k)$, then $X=M$ \\
    $X_i$ & The $i-$th factor, also, for another vectors, it is the $i-$th element\\
    $\hat{X}$ & Vector of the best estimation for input factors \\
    $\hat{X}_i$ & The value of best estimation for the $i-$th factor\\
    $\hat{X}_{\sim i}$ & Vector of the best estimation for all input factors but the $i-$th factor\\
    $\Omega$ & Model search space\\
    $\hat{\Omega}$ & Promissory model search space\\
    $\Theta$ & Confidence sub-contour box for $\hat{X}$ \\
    $M_{i:j,k:l}$ & Matricial notation to extract certain rows and columns for a given matrix $M$. It must be read as: \\
    {}& Select rows from $i$ to $j$ and columns from $k$ to $l$. $end$ represent the largest possible subindex\\
         \\ \hline
    \end{tabular}
    \end{table}

Suppose we have a model in the general form \eqref{eq:simplemodel} on which we want to perform both of the factor and CI estimation task over $k$ input factors $X=\set{X_1,X_2,...,X_k}$:

\begin{equation}\label{eq:simplemodel}
  Y=f(X ,t),\: t\in \tau, \: 
\end{equation}
Where $Y$ is the dynamic output, $f$ represents the deterministic model response function, $X$ is the vector of factors, and $\tau$ is the discretized domain of the model. As pointed out in the introduction, we define the search box following the logical sequence next: Being $X_i$ the $i-$th factor, then $X_i \in \left[\underline{I_i},\overline{I_i}\right]=I_i$, where $I_i$ is the factor interval and $\underline{I_i},\overline{I_i}$ represents its lower and upper bounds, respectively. Thus, the search box of the model is given by $\Omega =\set{I_1}\times...\times\set{I_k}$.

Once we define the search box, we need to select a function to assess the agreement between points in search space and real data (dissimilarity quantification), and then perform an optimization process. Since both model output and real data are vectors, a general approach to dissimilarity quantification is to define a loss function \cite{Seber2003}. It is possible to identify a general form for common loss functions introducing a parameter whose meaning is the importance of the distance between the output vector and the nominal one in a discrete point-to-point comparison sense \cite{Xiao2018}. A general loss function equation is given in \eqref{eq:error}, where $Y$ is the output vector obtained when simulating with a specific $X$, $\hat{Y}$ is the output vector associated to the real factors $\hat{X}$ (real data), $\alpha$ is the introduced distance-penalization factor, and $|\tau|$ is the number of discrete time elements in $\tau$.\\

\begin{equation}\label{eq:error}
Err(Y,\hat{Y})=\frac{1}{|\tau|}\sum_{t=1}^{|\tau|}{\left |{Y}_{t}-{\hat{Y}}_{t}
\right |^\alpha}{}
\end{equation}

\begin{figure}[htbp!]
    \centering
    \includegraphics[scale=0.5]{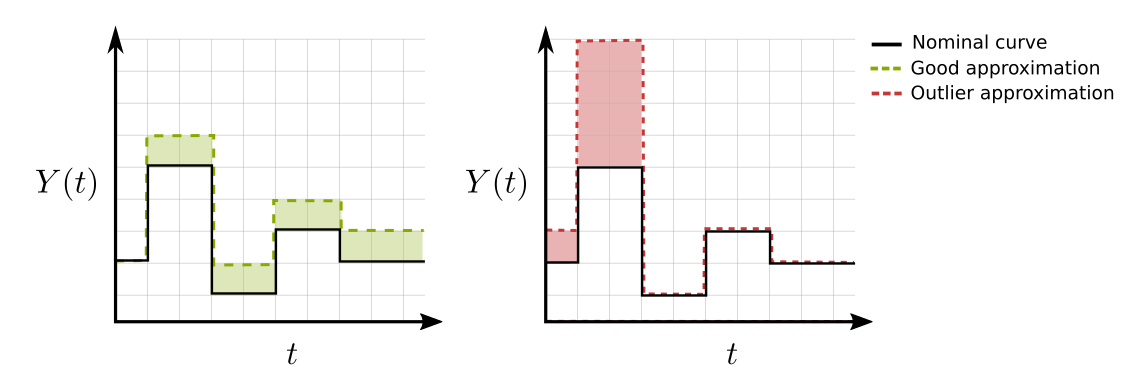}
    \caption{Example of dissimilarity quantification. Suppose we have a model with a given nominal output (black line) and we want to assess the fit of two different points in the search box (green and red lines) regard to the nominal output. From a graphic criterion, we may say that the green line has the best fit since it captures the behavior of the nominal model output over the whole model domain while the red line has the worst fit due to the high values it takes over the first interval of the model domain. However, if we apply the loss function MAE instead of MSE to assess the fit, we will get that both of the points in the search box have an equally good fit since the areas between black and colored curves are the same.  }
    \label{fig:outlier}
\end{figure}

The criteria for choosing a loss function must vary according to the study model and researchers goals. However, we recommend mean-squared error ($\alpha=2$) instead of mean-absolute error ($\alpha=1$) for most of the cases because of the outlier penalization of higher $\alpha$ values, as can be seen in Figure \ref{fig:outlier}. We support the above based on some principles of sensitivity analysis (SA) theory. SA states that the value of the most sensitive parameters almost determines the model output \cite{Saltelli2010, Saltelli2008, Sobol2001}. Therefore, the more similar are the values for the sensitive input factors, the more similar the model outputs should be. Then, as we map the search box into the least-squares surface generated by MSE, it would be easier to identify those factors that cause undesired model behaviors.

Thereby, for factor estimation, we define a new model with scalar output in the following form: 

\begin{align}
\begin{split}
    \phi(X,t)={}& Err(f(X,t),{\hat{Y}})
\end{split}\label{eq:scalarmodel}
\end{align}

Where $X$ is the vector to optimize, $Err$ is the loss function \eqref{eq:error} and $f(X.t)$ is the deterministic model \eqref{eq:simplemodel}. Thus, \eqref{eq:scalarmodel} is the quantification of dissimilarity between the real data (${\hat{Y}}$) and any other point ($X$) in the search space. Then, the factor estimation task has turned into an optimization problem in the form:

\begin{equation}\label{eq:min}
\begin{aligned}
& \underset{X\in \Omega}{\text{minimize}}
& & \phi(X,t)
\end{aligned}
\end{equation}

As mentioned in \cite{Draper1998}, if parameter combinations present several global or even local minima, the estimation algorithm may converge to unwanted or non-feasible stationary points of the search box. Thus, few factor estimation runs are not informative yet. It is highly recommended to perform an in-depth exploration of the search box through several factor estimations tasks starting optimization from different points, and then assess the results looking for the best estimations (including dissimilarity and other \textit{a priori} criteria) \cite{Seber2003}.

Once we get a promissory candidate from the factor estimation methodology ($\hat{X}$), we might proceed to the estimation of its CI. Following the general way to CI achievement, we should perform factor estimations and filter its results to get those that converge to $\hat{X}$. If we assume that minimum values of \eqref{eq:scalarmodel} almost determine $\hat{X}$, then we may estimate its CI by choosing the sets of factors that do not exceed in 10\% the minimum value we get in \eqref{eq:min} and applying them a robust measure of centrality, as \eqref{eq:cimedian}.

    \begin{equation} \label{eq:cimedian}
        X_i \in \left[median(X_i) \pm 1.96 \sqrt{\frac{\pi}{2}}\frac{\sigma_{i}}{\sqrt{n}}\right]
    \end{equation}
Where $n$ is the number of samples (filtered factor estimations), $\sigma_{i}$ the estimated standard deviation for factor $X_i$, and $median(X_i)$ is the median value of filtered values estimated for $X_i$. 

\begin{figure}[htbp!]
    \centering
    \includegraphics[scale=0.4]{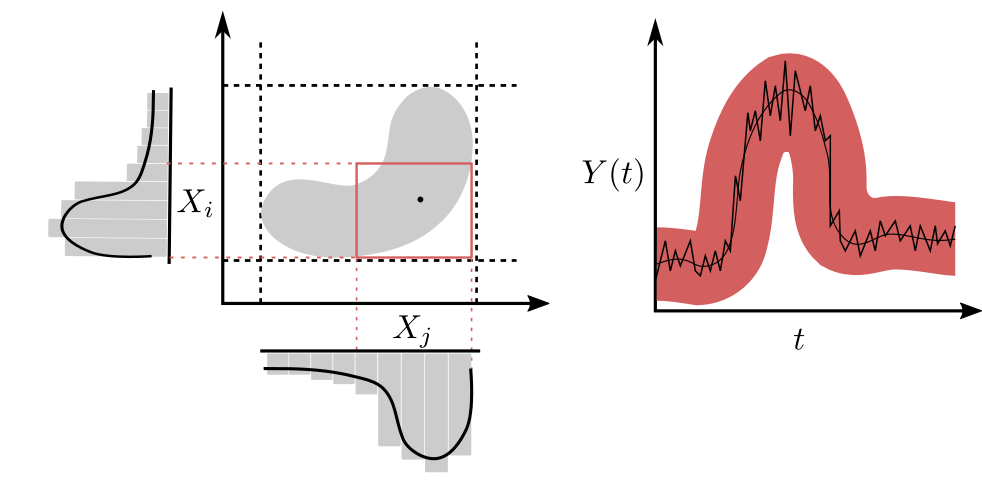}
    \caption{Here we have a model with dynamic output $Y(t)$ that depends on two input factors ($X_i$ and $X_j$). The confidence contour for some real data with uncertainty is the manifold inside the search box (left picture) such that, as we simulate each point in the manifold, we gradually clausure the data (right picture). We call to the clausure of the data as the confidence bands. Since each point within the manifold has a good fit with data, the axis on the left picture shows the probability distribution of factors $X_i$ and $X_j$ we would get from factor estimation task. Further, whether we traditionally estimate CI (dotted lines), then we get an overestimation of the confidence contour. Throughout the present work, we are interested in the estimation of the confidence sub-contour box (red box in the left picture) instead of the confidence contour or CI.}
    \label{fig:CSB}
\end{figure}

However, CI estimation has several troubles: as exposed previously, CI is an over-estimations of confidence contours (see for instance Figure \ref{fig:CSB}); its estimation depends on multiple optimization processes of \eqref{eq:min}; and the presence of global, or even local minima within the search box increases the CI uncertainty. As an alternative, it is still possible to map the whole search space into the loss function surface and then, identify the confidence contour for a given $\hat{Y}$. However, the confidence contour is a manifold that resides within search space and its achievement is challenging for large models \cite{Sandia2005}. Hence, we propose an alternative way to achieve a $n-$orthotope inside the confidence contour with some useful properties. We will call this $n-$orthotope as the CSB.

\section{\textbf{Construction of promissory search box for estimation of confidence sub-contour box}}
    From the previous section, we know that $\hat{X}$ is a global minimum candidate inside the confidence contour. To discard regions of the search box that do not belong to the confidence contour, we can perform an exploration of $\Omega$ through Monte-Carlo simulations drawing scatter-plots of \eqref{eq:scalarmodel} as a function of each parameter. However, since we are assessing dissimilarity for a given $\hat{X}$ instead of $\hat{Y}$, we will use the following redefinition of \eqref{eq:scalarmodel}:

\begin{align}
\begin{split}
    \phi(X,t)={}& Err(f(X,t),f(\hat{X},t))
\end{split}\label{eq:scalarexplorer}
\end{align}

Also, since $\hat{X}$ produces an error-free model output, we will delimit its confidence contour defining a threshold in the form:

\begin{align}
\begin{split}
    Threshold={}& Err(\lambda f(\hat{X},t),f(\hat{X},t))
\end{split}\label{eq:threshold}
\end{align}
where $\lambda$ represents the CSB uncertainty level. Thus, we say that an arbitrary set $X$ is outside of the confidence contour for $\hat{X}$ as long as $\phi(X,t)>Threshold$. On the other hand, from SA theory in \cite{Sobol1990}\cite{Saltelli2008} we can deduce that sensitive parameters mostly determine the model output and its non-feasible values should be related to the highest outputs of \eqref{eq:scalarexplorer}. Hence, drawing the scatter-plots of most sensitive factors we should be able to identify non-feasible subregions and eliminate them, shrinking the range for those factors. Once we shrink the sensitive factors interval, we can perform a new Monte-Carlo simulation and then, repeat the whole process until we achieve a $k-$orthotope where model outputs are almost within the confidence contour (confidence sub-contour box). Since we shrunk the interval of the most sensitive factors in each iteration, the sensitivity of other factors must take greater relevance; Hence, from the perspective of scatter plots, a correlation between factor value and model output must tend to zero. Thus, we also expect sensitivity indices to tend to be equal for most factors.\\

\begin{algorithm}[htbp!]
\DontPrintSemicolon
\KwData{$(\lambda,f,\hat{X},t,up,down,imax)$ such that $\lambda$ is the uncertainty level, $f$ is the objective function, $\hat{X}$ is the set of nominal parameters, $t$ is the simulation time, $up \in [1,2]$ is the scalar for upper accelerated search, $down \in [0,1]$ is the scalar for lower search, and $imax$ is the maximun number of iterations.}
\KwResult{$\hat{\Omega}$ as a matrix of $|\hat{X}|\times 2$, where each row represents a side of the $n-$orthotope (the promissory search box).}
\Begin{
$Threshold \longleftarrow Err\left(\lambda f\left(\hat{X},t\right),f\left(\hat{X},t\right)\right)$\;
$(\underline{\omega},\overline{\omega}) \longleftarrow (\emptyset,\emptyset)$\;
\For{$i \in \set{1,2,...,|\hat{X}|}$}{
$(\tau,\gamma,\delta,c) \longleftarrow (1,up,up,0)$\;
\While{$c<2$}
{$Y \longleftarrow Err\left(f\left(\set{\hat{X}_{\sim i},\delta\hat{X}_i},t\right),f\left(\hat{X},t\right)\right)$\;
\eIf{$Y<Threshold$}
{$\delta \longleftarrow \gamma^{\tau+1}$\;}
{\eIf{$Y>Threshold\times1.1$}{$\omega \longleftarrow Bisection(f,\gamma,\tau,\hat{X}_{\sim i},\hat{X}_i,Threshold,imax)$\Comment{See Algorithm \ref{al:bisection}}\;}
{$\omega \longleftarrow \delta\hat{X}_i$\;

}\eIf{$\gamma = up$}{$\overline{\omega_i} \longleftarrow \omega$}{$\underline{\omega_i} \longleftarrow \omega$}
$(\tau,\gamma,\delta,c) \longleftarrow (0,down,down,c+1)$\;}
$\tau \longleftarrow \tau + 1$\;
\If{$\tau > imax$}{$\omega \longleftarrow \delta\hat{X}_i$\;
\eIf{$\gamma = up$}{$\overline{\omega_i} \longleftarrow \omega$}{$\underline{\omega_i} \longleftarrow \omega$}
$(\tau,\gamma,\delta,c) \longleftarrow (1,down,down,c+1)$\;
}}
}$\hat{\Omega} \longleftarrow \set{\underline{\omega},\overline{\omega}}$\;
\KwRet{$\hat{\Omega}$}
}
\caption{Once-At-Time promissory search box\label{al:oat}}
\end{algorithm}

However, there are still remarkable troubles that arise when applying the above methodology: Since the model search box ($\Omega$) is usually much larger than the contour confidence and non-linear models have multiple local (or even global) minima, then localization of non-feasible regions must be unlikely, unless we have a complete exploration of $\Omega$, i.e., as the number of samples $N$ tends to infinity.

To avoid the troubles exposed above, we propose the construction of a promissory search box in Algorithms \ref{al:oat} and \ref{al:bisection}. It is possible whether we take $\hat{X}$ as a global minimum: For each factor $X_i$, we take $\hat{X}_i$ as a starting value. Then, keeping all the other factors fixed ($\hat{X}_{\sim i}$), we proceed to increase and decrease the $\hat{X}_i$ value until the model output falls into the border of the confidence contour. Thus, we can find a new interval for each factor that defines a new promissory search space $\hat{\Omega}$. As we use the contribution of each factor alone to reach the bound of the confidence contour, we are also neglecting the contribution of interactions. Thus, there should be regions related with extreme values of some factors beyond the confidence contour, i.e., we might expect that as $X_i$ take values closest to its interval bounds in $\hat{\Omega}$, the model output takes higher values. It is feasible that some local minima remain within $\hat{\Omega}$, nevertheless, taking priority over $\hat{X}$ it may be possible to discard those local minima related to specific combinations of factors.\\

 \begin{algorithm}[htbp!]
\DontPrintSemicolon
\KwData{$(f,\gamma,\tau,\hat{X}_{\sim i},\hat{X}_i,Threshold,imax)$ such that $f$ is the objective function, $\gamma$ is the scalar for upper or lower search from Algorithm \ref{al:oat}, $\tau$ is the current search step, $\hat{X}_{\sim i}$ is the vector of the nominal factors but the $i-$th one, $\hat{X}_i$ is the $i-$th nominal factor, $Threshold$ is the limit for uncertainty level, and $imax$ is the maximum number of iterations allowed.}
\KwResult{$\omega$ as the desired value for parameter $X_i$.}
\Begin{
$a \longleftarrow \gamma^{\tau}$\;
$b \longleftarrow \gamma^{\tau+1}$\;
$\delta \longleftarrow \frac{a+b}{2}$\;
\While{True}{$Y \longleftarrow Err\left(f\left(\set{\hat{X}_{\sim i},\delta\hat{X}_i},t\right),f\left(\hat{X},t\right)\right)$\;
\eIf{$Threshold < Y < Threshold\times 1.1$}{$\omega \longleftarrow \delta\hat{X}_i$\;
\KwRet{$\omega$}}{\eIf{$Y<Threshold$}{$a \longleftarrow \delta$\;
$\delta \longleftarrow \frac{a+b}{2}$}{$b \longleftarrow \delta$\;
$\delta \longleftarrow \frac{a+b}{2}$}}
$\tau \longleftarrow \tau+1$\;
\If{$\tau >imax$}{$\omega \longleftarrow \delta\hat{X}_i$\;
\KwRet{$\omega$}}}
}
\caption{Bisection\label{al:bisection}}
\end{algorithm}

\section{\textbf{Uncertainty-based confidence interval shrink (algorithm)}}
 Following the idea from the previous section for CSB estimation, in this section we will present four algorithms: Monte-Carlo simulations with Latin-hypercube sampling (Algorithm \ref{al:MC}), uncertainty-based factor interval shrink (Algorithm \ref{al:UCI}), protection of $\hat{X}$ (Algorithm \ref{al:PC}), and change uncertainty-based shrink parameterization (Algorithm \ref{al:CP}). Also, we will expose the Latin-hypercube sampling and the uncertainty-based interval shrink procedure applying histograms.\\
 
 \begin{algorithm}[htbp!]
\DontPrintSemicolon
\KwData{($f,\hat{X},\hat{\Omega},N$) such that $f$ is the objective functon, $\hat{X}$ is the vector of nominal factors, $\hat{\Omega}$ is the promissory search box, and $N$ is the sample size.}
\KwResult{($Y,M$) where $Y$ is the vector of sorted outputs after Monte Carlo simulation and $M$ is the design matrix for the Monte Carlo simulation}
\emph{sort is a function that receives a vector $[y_1,y_2,...,y_n]$ and returns it sorted alongs the sorting order: $[a,b,...z] \mid y_a\leq y_b \leq ... \leq y_z$}\;
\Begin{
$M \longleftarrow latinHypercube(\hat{\Omega},N)$\;
\For{$i\in \set{1,2,...,N}$}
{$y_i \longleftarrow Err(f(M_{i,1:end},t),f(\hat{X},t))$\;}
$(Y,order) \longleftarrow sort([y_1,y_2,...,y_n])$\;
$M \longleftarrow M_{1:end,order}$ \Comment{sorted design matrix according to $Y$}\;
\KwRet{$(Y,M)$} 
}
\caption{Monte Carlo simulation\label{al:MC}}
\end{algorithm}

  \begin{algorithm}[htbp!]
\DontPrintSemicolon
\KwData{$\left(\lambda,f,\hat{\Omega},\hat{X},N,imax,\eta,\xi, \Delta \right)$ such that $\lambda$ is the uncertainty level (same as Algorithm \ref{al:oat}), $f$ is the objective function, $\hat{\Omega}$ is the promissory search space from Algorithm \ref{al:oat}, $\hat{X}$ is the vector of nominal factors, $N$ is the sample size for each iteration, $imax$ is the maximum number of iterations, $\eta$ is the number of higher outputs eliminated per iteration, $\xi$ is the threshold for histogram-based cut (see Figure \ref{fig:firstIteration}), and $\Delta$ is the stop criterion (\%).}
\KwResult{$\Theta$ as a matrix of $\left|\hat{X}\right|\times 2$, where each row represents a side of the hypercube (The Confidence Sub-contour Box).}
\Begin{
$Threshold \longleftarrow Err\left(\lambda f\left(\hat{X},t\right),f\left(\hat{X},t\right)\right)$\;
$\eta \longleftarrow round(\eta\times N)$\;
$c \longleftarrow 1$\;
$\Theta \longleftarrow \hat{\Omega}$\;
\While{True}{
$(Y,M) \longleftarrow MonteCarlo(f,\hat{X},\Theta,t,N)$\Comment{See Algorithm \ref{al:MC}}\;
$\epsilon \longleftarrow \sum_{i=1}^N(Y_i\leq Threshold)$\;
\If{$\epsilon/N\geq \Delta$}{\KwRet{$\Theta$}}
$\psi \longleftarrow max(\epsilon,\eta)$\;
$sM \longleftarrow M_{1:\psi,1:end}$\;
$nBins \longleftarrow \set{x\in \field{N}: (x \text{ divide a } N) \land (x\leq N/10) }$\;
$\tau \longleftarrow 1$\;
$recort \longleftarrow \text{True}$\;
\While{$recort$}{
\For{$i \in \set{1,2,...,|\hat{X}|}$}{
$\left(\underline{\theta_i},\overline{\theta_i}\right) \longleftarrow CSBhistogram(sM_{1:end,i},{nBins}_\tau,\xi,\Theta_{i})$\Comment{${nBins}_\tau$ is the $\tau -$th element of $nBins$}\;
\If{$\left(\underline{\theta_i}>\hat{X}_i\right)\lor \left(\overline{\theta_i}<\hat{X}_i\right)$}{
$\left(\underline{\theta_i},\overline{\theta_i}\right) \longleftarrow protectCriteria\left(\underline{\theta_i},\overline{\theta_i},\hat{X}_i\right)$\Comment{See Algorithm \ref{al:PC}}\;
}
}
$\Phi \longleftarrow \set{\underline{\theta},\overline{\theta}}$\;
\eIf{$\Theta = \Phi$}{$(\tau,sM,\psi) \longleftarrow changeParameters(\psi,\tau,nBins,M,N)$\Comment{See Algorithm \ref{al:CP}}}{$\Theta \longleftarrow \Phi$\;
$recort \longleftarrow \text{False}$}}
$c \longleftarrow c+1$\;
\If{$c>imax$}{\KwRet{$\Theta$}}
}
}
\caption{Uncertainty-based interval shrink\label{al:UCI}}
\end{algorithm}

 To make easier the presentation of algorithms, we will introduce first the original and most simple form of the Latin-hypercube sampling plan for Monte-Carlo simulation purposes \cite{Olsson2003}:
 Let $N$ denote the required number of samples and $k$ the number of factors. A $N\times k$ matrix \textbf{P}, in which each of the $k$ columns is a random permutation of $1,2,...,N$, and a $N\times k$ matrix \textbf{R} of independent random numbers $\sim U(0,1)$ are established. These matrices form the sampling plan, representing by \textbf{S} as:
 \begin{equation}\label{eq:lhs}
     \textbf{S}=\frac{1}{N}(\textbf{P}-\textbf{R})
 \end{equation}
 Each element of \textbf{S}, $S_{i,j}$ is then mapped according to its target marginal distribution as:
 \begin{equation}\label{eq:lhs2}
     x_{i,j}=F^{-1}_{x_j}(S_{i,j})
 \end{equation}
 Where $F^{-1}_{x_j}$ is the inverse of the target cumulative distribution function for variable $j$. Thus, we define the design matrix $M$ as:
 \begin{equation}
     M=\left [\begin{matrix*}[c]
& x_{i,1}&\hdots &x_{i,k} \\
& \vdots&\vdots &\vdots \\
& x_{N,1}&\hdots &x_{N,k} \\
\end{matrix*}
\right ]
 \end{equation}
 
 Coming back to the main idea of CSB achievement, once we get $\hat{X}$, we perform an uncertainty analysis through Monte-Carlo simulations to map $\hat{X}$ into the loss-function surface \eqref{eq:scalarexplorer}. We expect non-feasible values for some of the most sensitive parameters to cause the model outputs with the highest dissimilarity. Since those non-feasible values should proceed from closer samples to the interval bounds, thereby we propose the following schema to achieve the $\hat{X}$ CSB at $\lambda$ uncertainty level:

   \begin{figure}[htbp!]
    \centering
    \includegraphics[scale=0.4]{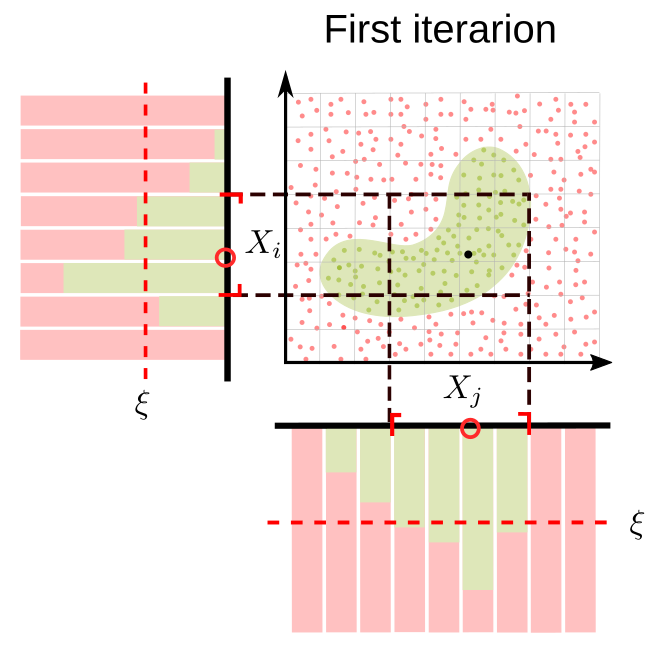}
    \caption{Graphical representation of CSBhistogram function for a model that depends on two factors $X_i$ and $X_j$. The graph axes represent the promissory search box $\hat{\Omega}$, the red circles in the axes are the nominal factor values ($\hat{X}_i$, $\hat{X}_j$), and the black dot is a promissory point ($\hat{X}$) from factor estimation task on which we want to estimate the CSB. For the CSB estimation, we perform an uncertainty analysis using a Latin-hypercube sampling; thereby, we guarantee that each column or row of the grid we have plotted in $\hat{\Omega}$ has the same number of samples. Then we classify the samples as they overcome (red dots) or not (green dots) a previously defined threshold of dissimilarity and plot them for each factor as a histogram. From SA theory, we know that non-feasible values for the most sensitive parameters correlate with strange model outputs. Hence, we develop a shrink criterion based on the form of the histogram: the new CSB candidate for next iteration is the promissory box resultant from the exclusion of those grid rows that have less frequency than $\xi\%$ of the larger bin in the green histogram. The CSB is fully estimated when at least $95\%$ of the samples are below the threshold.}
    \label{fig:firstIteration}
\end{figure}

 \begin{enumerate}
     \item Perform sampling of $\hat{\Omega}$ where each factor interval is divided into $N$ sub-intervals (same $N$ as the total number of samples) of equal length, then, the factor is sampled taking a value from each sub-interval in combination with random permutations of others factors intervals (Latin-hypercube sampling). As a consequence of this sampling approach, if we split the interval of any factor into $r$ sub-intervals of equal length and $r$ is any divisor of $N$, then we will have an equal number of samples in each sub-interval.
     \item Simulate the model with the $N$ samples from the previous step and sort the outputs from lower to higher values. See Algorithm \ref{al:MC}.
     \item Plot a histogram of $r$ bins for each factor, where $r$ is a divisor of $N$ such that the number of samples in every bin (given by $samples=\frac{N}{r}$) overcomes a defined number. This number would be enough to capture information about factor contribution and possible interactions. Since each bin has the same number of samples, histogram must be uniform (see Figure \ref{fig:firstIteration}).
     \item Remove $\eta\%$ of samples linked to the higher model output values from the histogram. From sensitivity analysis theory, the histogram shape for most sensitive factors must change dramatically, especially for extreme interval values. For less sensitive factors, histogram shape must remain as uniform. See Figure \ref{fig:firstIteration}.
     \item Remove bins whose number of samples is less than $\xi\%$ of the bin with the most of samples. Then, define the new factor interval with the minimum and the maximum factor value that belongs to some bin.
     \item If it is not possible to perform any factor interval shrink, then change $r$, $\xi$ or eliminate more samples linked to higher output values. See Algorithm \ref{al:CP}.
     \item If the procedure above attempts to eliminate some $\hat{X}_i$, then revert the procedure or shrink the $i-$th interval forcing the method to keep $\hat{X}_i$. See Algorithm \ref{al:PC}.
     \item Repeat the above steps until $\Delta\%$ of model outputs within the confidence contour.
 \end{enumerate}
 
  \begin{algorithm}[htbp!]
\DontPrintSemicolon
\KwData{($\underline{\theta_i},\overline{\theta_i},\hat{X}_i$) such that $\underline{\theta_i}$ is the current inferior bound for the $i-$th factor from Algorithm \ref{al:UCI}, $\overline{\theta_i}$ is the current superior bound for the $i-$th factor from Algorithm \ref{al:UCI}, and $\hat{X}_i$ is the nominal value for the $i-$th factor.}
\KwResult{($\underline{\theta_i},\overline{\theta_i}$) where $\underline{\theta_i}$ is an appropriate inferior bound and $\overline{\theta_i}$ is an appropriate superor bound for the $i-$th factor.}
\Begin{
\eIf{$\underline{\theta_i}>\hat{X}_i$}{$(\underline{\theta_i},\overline{\theta_i}) \longleftarrow \left(\hat{X}_i -(\hat{X}_i+\overline{\theta_i})/2\times 0.1,\overline{\theta_i}-(\hat{X}_i+\overline{\theta_i})/2\times 0.1\right)$
}
{$(\underline{\theta_i},\overline{\theta_i}) \longleftarrow \left(\underline{\theta_i} +(\hat{X}_i+\underline{\theta_i})/2\times 0.1,\hat{X}_i+(\hat{X}_i+\underline{\theta_i})/2\times 0.1\right)$}
\KwRet{$(\underline{\theta_i},\overline{\theta_i})$}
}
\caption{protectCriteria\label{al:PC}}
\end{algorithm}
 
 \begin{algorithm}[htbp!]
\DontPrintSemicolon
\KwData{($\psi,\tau,nBins,M,N$) such that $\psi$ is the current number of higher outputs to eliminate from Algorithm \ref{al:UCI}, $\tau$ is the current element of $nBins$ that determine the grid for $CSBhistogram$ (Figure \ref{fig:firstIteration}), $nBins$ is the vector of $N$ divisors, $M$ is the  design matrix for the current iteration in Algorithm \ref{al:UCI}, and $N$ is the sample size.}
\KwResult{($\tau,sM,\psi$) As the new parameterization for current Algorithm \ref{al:UCI} iteration.}
\Begin{
\eIf{$\tau+1 \leq |nBins|$}{$\tau \longleftarrow \tau +1$}{$\tau \longleftarrow 1$\;
$\psi \longleftarrow round((\psi/N)^{1.1})\times N$}
$sM \longleftarrow M_{1:\psi,1:end}$\;
\KwRet{$(\tau,sM,\psi)$}
}
\caption{changeParameters\label{al:CP}}
\end{algorithm}
 
Step 4 from the previous schema suggests removing $\eta\%$ of samples from higher model outputs, though we implemented an additional feature in Algorithm \ref{al:UCI}: the removal of $\eta\%$ until $\epsilon\%, \eta < \epsilon$ of model outputs within the confidence contour, thenceforth, remove the $\epsilon\%$. We did this trying to give major precision to the algorithm.

\section{Algorithm application case using GSUA-CSB toolbox}\label{sec:appcase}

To test Algorithms \ref{al:oat} to \ref{al:CP} we implemented and perform both of proposed and traditional routines for factor estimation and confidence intervals estimation (CSB and CI) fitting an epidemiological model \eqref{DengueModel} to real data. We managed to validate the stated CSB properties and exposed new ones through validation with sensitivity and uncertainty analysis. All the routines we followed are implemented in GSUA-CSB toolbox \cite{gsua}, free available in \url{https://www.mathworks.com/matlabcentral/fileexchange/72637-gsua-csb}.

\subsection{Methods}

\subsubsection{Mathematical model}
 In this section, we will introduce the model implemented to estimate parameters and its confidence intervals for the dengue outbreak produced in a municipality of Neiva (Colombia) during 2011-2012. We modified a general vector-borne disease model, proposed in \citep{Tuncer2018} for describing Zika dynamics, which represents disease transmission in human and mosquitoes populations (see (\ref{DengueModel})). It is possible to use (\ref{DengueModel}) to describe dengue basic dynamic, because the same genus mosquitoes may transmit dengue and Zika.
    
    \begin{equation}\label{DengueModel}
        \begin{array}{lcl}
            \dot{M}_s = \Lambda_v - \frac{\beta_m H_i M_s}{H} - \mu_m M_s \\
            \dot{M}_i = \frac{\beta_m H_i M_s}{H} - \mu_m M_i  \\
            \dot{H}_s = \mu_h H - \frac{\beta_h M_i H_s}{M} - \mu_h H_s \\
            \dot{H}_i = \frac{\beta_h M_i H_s}{M} - (\mu_h + \gamma_h)H_i \\
            \dot{H}_r = \gamma_h H_i - \mu_h H_r
        \end{array}
    \end{equation}
    
  For (\ref{DengueModel}) we assume a total mosquito population variable in time, where $M= M_s+M_i$, and a total human population constant as $H=H_s+H_i+H_r\approx const$. Also, there are no recovered individuals that lose immunity and neither horizontal transmission. We list the depended variables and parameters of the model in Table (\ref{Tab:definition})
    
   \begin{table}[htbp]
   \centering
   \caption{Definition of model state variables and parameters\label{Tab:definition}}
    \begin{tabular}{cl}
    \hline
    \textbf{Factors} & \multicolumn{1}{c}{\textbf{Definition}} \\ \hline
    $M_s$           & Number of susceptible mosquitoes        \\
    $M_i$           & Number of infected mosquitoes           \\
    $M$             & Total number of mosquitoes              \\
    $H_s$           & Number of susceptible humans            \\
    $H_i$           & Number of infected humans               \\
    $H_r$           & Number of recovered humans              \\
    $H$             & Total number of humans                  \\
    $\Lambda_v$     & Mosquitoes population recruitment rate  \\
    $\beta_m$       & Transmission rate human-mosquito        \\
    $\mu_m$         & Mosquito mortality rate                 \\
    $\beta_h$       & Transmission rate mosquito-human        \\
    $\mu_h$         & Human mortality rate                    \\
    $\gamma_h$      & Infected human recovery rate            \\ \hline
    \end{tabular}
    \end{table}

Also, we chose MSE instead of MAE as a loss function for model \ref{DengueModel} basing our decision on the following consideration:\\
Models of Vector-borne diseases are commonly inferior-bounded since negative outputs for population sizes are senseless, and further, outputs of infected hosts and vectors represent a small percentage of its respective total population sizes for the study zone. Thus, strangest output behaviors should belong to some domain $[t_a,t_b]$ such that ${Y}_{i}>>{\hat{Y}}_i, \: \forall i \in [t_a,t_b]$. Hence, a loss function with a high value for $\alpha$ really must improve the detection of the strangest curves penalizing said high output values.

\subsubsection{Dengue cases data}
   The municipality of Neiva has its location in the valley of Magdalena between the central and eastern mountain ranges at 442 masl. Its main temperature is 27{\grad}C with an average annual total rainfall of 1346mm. These conditions are favorable to the \textit{Aedes aegypti} reproduction and maintenance \cite{Yang2011, Takahashi2005}, which is related to endemic dengue cases with occasional epidemic outbreaks. For this study, we implemented the number of dengue cases for the municipality of Neiva during 2011 and 2012, where the data begins in the 49th epidemiological week in 2011 and ends in the 49th epidemiological week in 2012 (53 weeks). Neiva health secretariat and SIVIGILA software provided the reported cases of dengue \cite{SIVIGILA2010}.

\subsubsection{Estimation of factors and traditional confidence intervals for dengue model \label{M:fitting}}
    To identify the proper parameters sets which $H_i$ in (\ref{DengueModel}) fit with real dengue cases, we perform several single-objective factor estimations implementing the function \textit{gsua\_pe} from GSUA-CSB toolbox with the following tools and methods: we numerically solved the model using the ode45 function which implements the Dormand-Prince variable-step method \cite{Shampine1997}, and the function \textit{lsqcurvefit} to solve the nonlinear curve-fitting problem with a function tolerance of $1*10^{-6}$, randomly selecting the starting point from the model search box defined in Table \ref{Table:Results}. To calculate traditional CI, we apply the filtering strategy and the median estimator exposed in Section \ref{sec:background}.
    
\subsubsection{CSB algorithms parameterization in GSUA-CSB toolbox}

We estimated multiple times CSB for the same $\hat{X}$ using the functions \textit{gsua\_oatr} and \textit{gsua\_uci} (Algorithms \ref{al:oat} and \ref{al:UCI} respectively) implemented in GSUA-CSB toolbox. For dengue disease, there is high uncertainty in real data that not only comes from measurement but also diagnosis itself \citep{Toan2015}\citep{Duong2015}. Hence, we chose an uncertainty level of $\lambda=30\%$. We set parameters $up$, $down$ and $imax$ for Algorithm \ref{al:oat} at his default values: 1.5, 0.7, and 100, respectively. For Algorithm \ref{al:UCI} we chose a sample size per iteration of $N=1000$, a number of maximum iterations of $imax=500$, a removal of higher model outputs of $\eta =50\%$, and a confidence level of $\Delta=95\%$.

\subsubsection{Sensitivity analysis routines}

The sensitivity analysis (SA) aims to quantify how uncertainty in the output of a model can be apportioned to the uncertainty of the input factors \cite{Saltelli2010}\cite{Archer1997}. We performed SA to assess one of the theoretical properties stated for CSB: since we shrunk the interval of most sensitive factors, all factors must tend to be equally relevant for the model, i.e., all factors must tend to have the same sensitivity index. For SA procedures, we chose Saltelli indices proposed in \cite{Saltelli2010} and implemented in \textit{gsua\_sa} function from GSUA-CSB toolbox with default parameterization.

On the other hand, to guarantee informative SA, we perform a convergence analysis of the indices. The convergence analysis consists of performing several SA increasing the sample size until sensitivity indices stop varying. Since \textit{gsua\_sa} function provides both of first-order and total-order sensitivity indices, we implement three graphical criteria to assess convergence: convergence of each total-order index alone, the convergence of summation of total-order index, and convergence in the sum of first-order sensitivity indices regard the sum of absolute values for the first-order indices. It is possible to implement the last criterion since the Saltelli estimator for first-order sensitivity indices allows negative values, but, from the definition, sensitivity indices might not be negative.

Saltelli indices implemented in GSUA-CSB toolbox are estimated on the MSE surface instead of the vectorial model output, i.e., they are scalar indices for dissimilarity contribution quantification. Hence, it is necessary to provide both $\hat{X}$ and $M$, where $M$ is a design matrix as presented for Monte-Carlo simulation. Design matrix $M$ was generated from the estimation ranges in Table \ref{Table:Results}. Thereby, we perform convergence analysis for the search box, i.e., the selected sample size must provide informative SA for any subspace in the search box.

\subsection{Results and discussion}

\subsubsection{Traditional confidence intervals versus confidence sub-contour box}
\begin{figure}[htbp!]
        \centering
        \includegraphics[width=8.5cm]{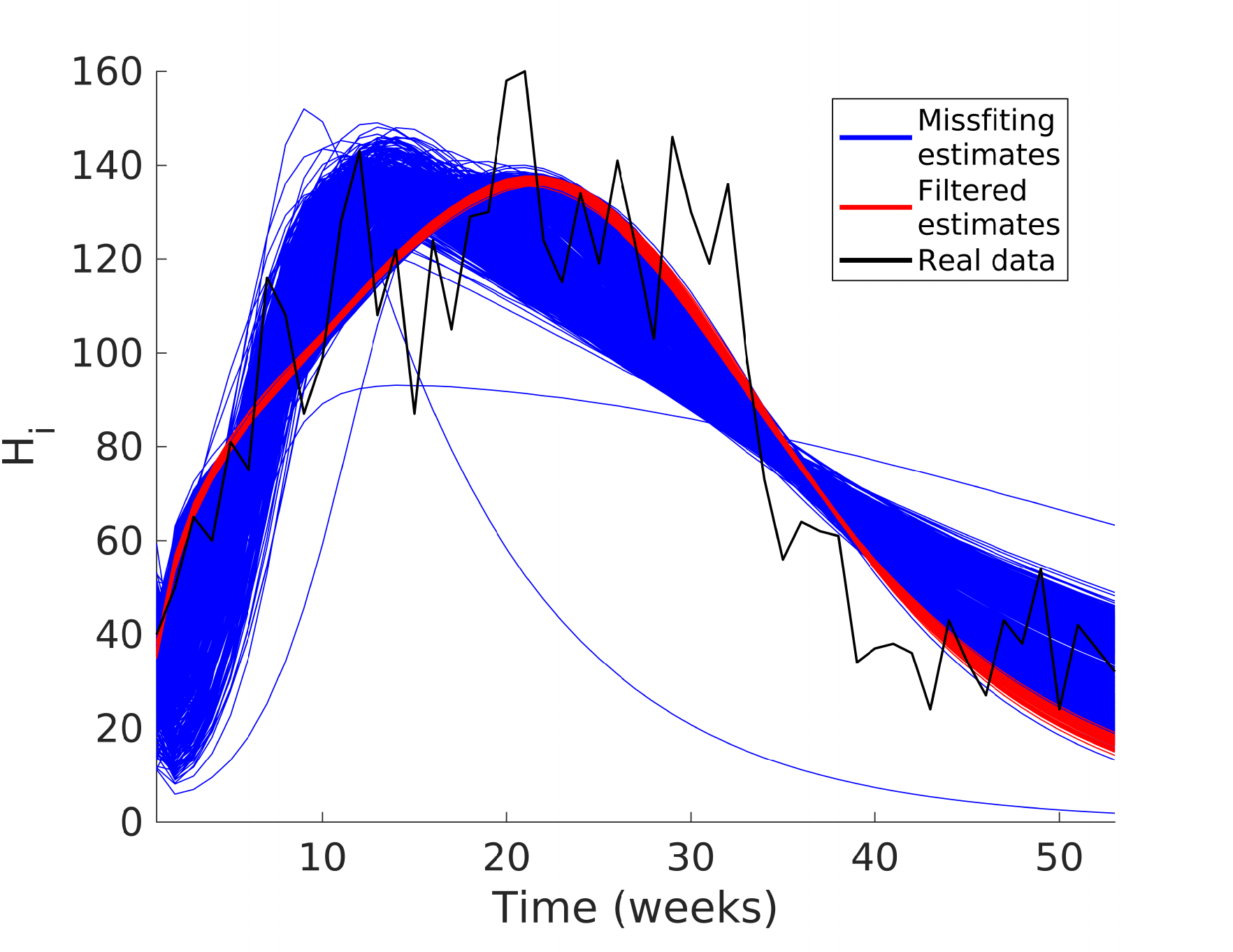}
        \caption{Simulations of over 14000 factor estimations; where the red lines correspond to the 3990 best-fitted curves and blue ones correspond to the misfitted remaining. Some curves were not included because they moved so far away from the real data that it was no longer possible to observe them.}
        \label{fig:Estimation}
    \end{figure}

For estimation of traditional CI, we performed 14116 factor estimation tasks on the model (\ref{DengueModel}), and then we applied the filtering criteria described in Section \ref{M:fitting} to avoid local minima. Thus, we got 3991sets of factors that did not overcome 10\% of the lowest minimum got (52.8 in MSE), as seen in Figure \ref{fig:Estimation}. Note that the filtered sets follow the same trend as real data, unlike the remaining curves. Then, we estimated the nominal curve $\hat{X}$ through the statistical median for the ten factors, as well as its traditional CI. We also perform 100 CSB estimations and present the results for a random one, as seen in Table \ref{Table:Results} along with biological ranges for each factor.

\begin{figure}
        \centering
        \begin{subfigure}[ht]{0.47\textwidth}
            \includegraphics[width=1\textwidth]{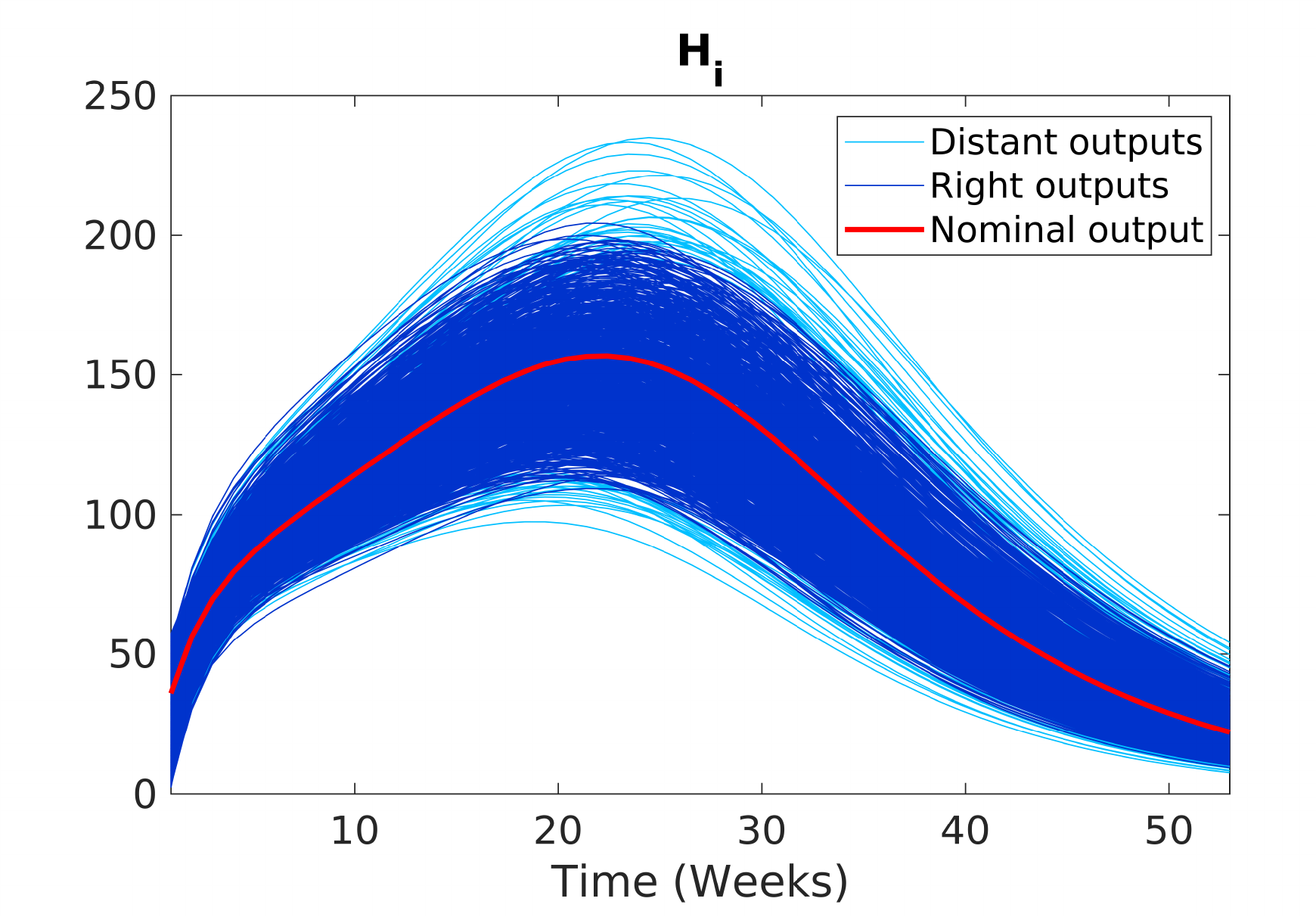}
            \caption{}\label{fig:ua_csb}
        \end{subfigure}
        \hfill
        \centering
        \begin{subfigure}[ht]{0.47\textwidth}
            \includegraphics[width=1\textwidth]{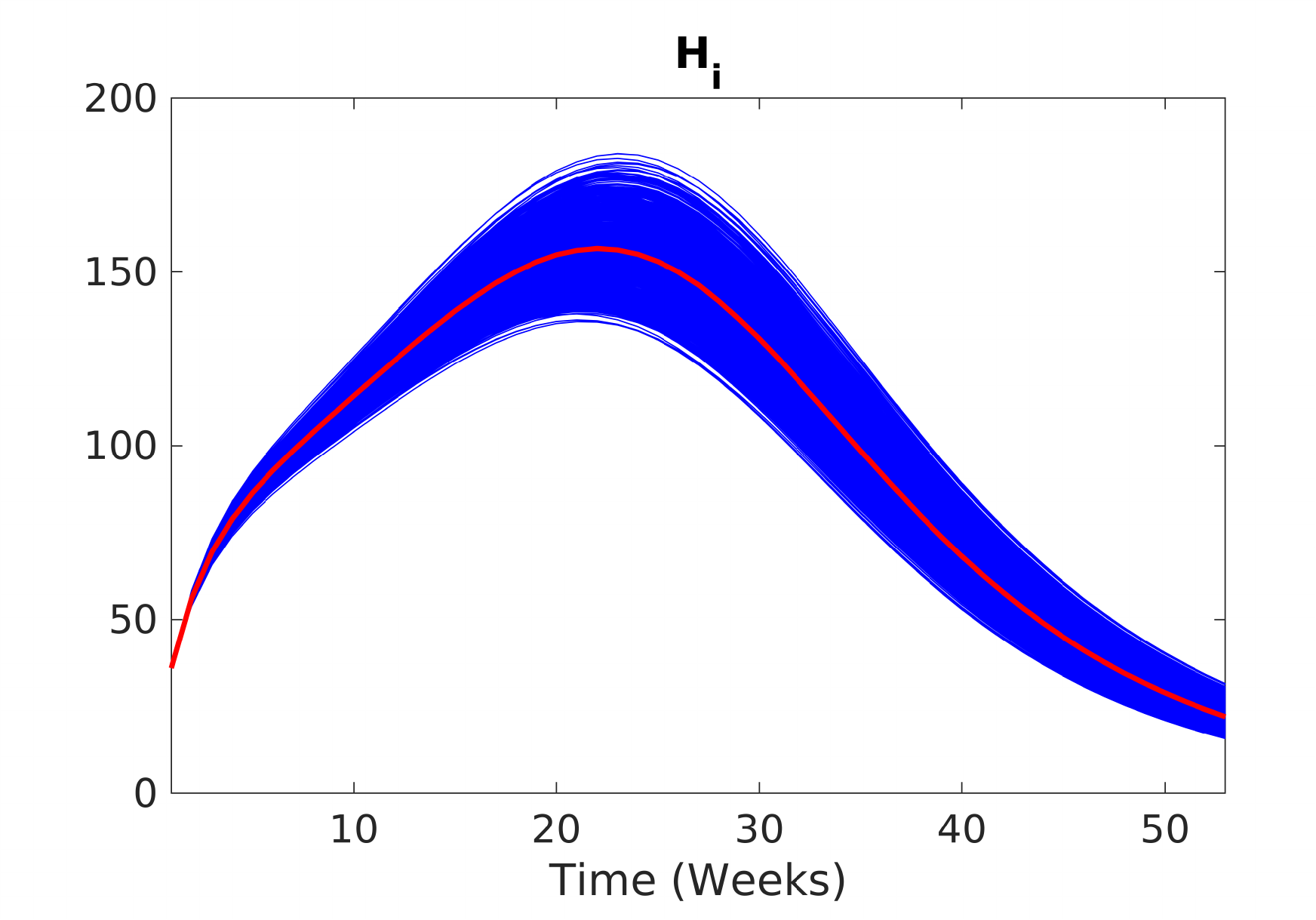}
            \caption{}\label{fig:ua_ci}
        \end{subfigure}
        \caption{Uncertainty analysis for both of confidence intervals presented in Table \ref{Table:Results} with 1000 Monte-Carlo samples. a) Uncertainty analysis of estimated CSB. b) Uncertainty analysis of traditonal median CI. The nominal output is given in red color while those outputs that overcomes the defined uncertainty level threshold $\Lambda=30\%$ are given in light blue color.}
        \label{fig:ua}
    \end{figure}

From Table \ref{Table:Results} we note that median CI are subsets of median CSB, except for $\beta_m$ and $\beta_h$, where superior bounds for both intervals almost coincide. Another remarkable result is given in Figure \ref{fig:ua}, where we perform an uncertainty analysis for both confidence intervals. Note that although uncertainty analysis for median CI is a subregion of uncertainty analysis for median CSB, median CI is a robust estimator where the length of the intervals depends on the number of factor estimations we perform, while for median CSB we can define the uncertainty level we want.

Computational cost is another important topic when looking for methods to estimate confidence intervals. For instance, we required 14116 factor estimation processes for model \eqref{DengueModel} to achieve enough information for median CI estimation. Each factor estimation process requires up to 1000 set of factors evaluations (and requires more than the model gets larger), then we performed about 14.1 million of model evaluations. On the other hand, a regular CSB estimation process requires from 100 up to 200 iterations, hence, it represents a cost of 0.1 to 0.2 million of model evaluations.

               \begin{table}[htbp]
               \centering
               
        \caption{List of the factors estimated for model \eqref{DengueModel}, its estimation ranges (search box) and nominal values ($\hat{X}$). Also, we show both of confidence intervals for the dengue model \eqref{DengueModel}, using median distribution (traditional CI) and a regular CSB estimation. \label{Table:Results}}
        \begin{tabular}{ccccc}
        \hline
        \textbf{Factors} & \textbf{\begin{tabular}[c]{@{}c@{}}Estimation \\ ranges\end{tabular}} & \textbf{\begin{tabular}[c]{@{}c@{}}Nominal\\ values\end{tabular}} & \textbf{\begin{tabular}[c]{@{}c@{}}Median\\ CI\end{tabular}} & \textbf{CSB} \\ \hline
        $M_s(0)$         & 0 - 20000000                                                          & 2110000                                                           & 2070000 - 2160000                                                   & 2070000 - 2660000   \\
        $M_i(0)$         & 0 - 1000                                                              & 670                                                               & 660 -   680                                                         & 650 - 800           \\
        $H_s(0)$         & 150000 - 400000                                                       & 281000                                                            & 279000 -    284000                                                  & 272000 -    335000  \\
        $\Lambda_v$      & 0 - 20000                                                             & 7800                                                              & 7600 - 7900                                                         & 7400 -  12000       \\
        $\beta_m$        & 0 - 4                                                                 & 0.064                                                             & 0.059 - 0.068                                                       & 0.054 - 0.066       \\
        $\mu_m$          & 0 - 0.9                                                               & 0.1665                                                            & 0.1664 - 0.1665                                                     & 0.153   - 0.169     \\
        $\beta_h$        & 0 - 4                                                                 & 0.48                                                              & 0.46 - 0.50                                                         & 0.44 -  0.49        \\
        $\mu_h$          & 0 - 0.0009                                                            & 0.00066                                                           & 0.00065 - 0.00067                                                   & 0.00007 -   0.00150 \\
        $\gamma_h$       & 0.5 - 1.8                                                             & 0.500                                                             & 0.497 - 0.503                                                       & 0.466 - 0.518       \\ \hline
        \end{tabular}
        \end{table}

\subsubsection{Uniqueness of the confidence sub-contour box}

We have assumed that the CSB estimation determines a unique $n-$orthotope within the confidence contour. We support the above since each iteration of intervals shrinking is lead by an approximation equivalent to sensitivity indices estimation from scatter plots. However, we have noticed that, as the contribution of factors is decreased by the shrink of its interval, the algorithm likely shrink some intervals on a stochastic way instead of the sensitivity-led way.

\begin{figure}[htbp!]

        \centering
        \begin{subfigure}[ht]{0.8\textwidth}
            \includegraphics[width=1\textwidth]{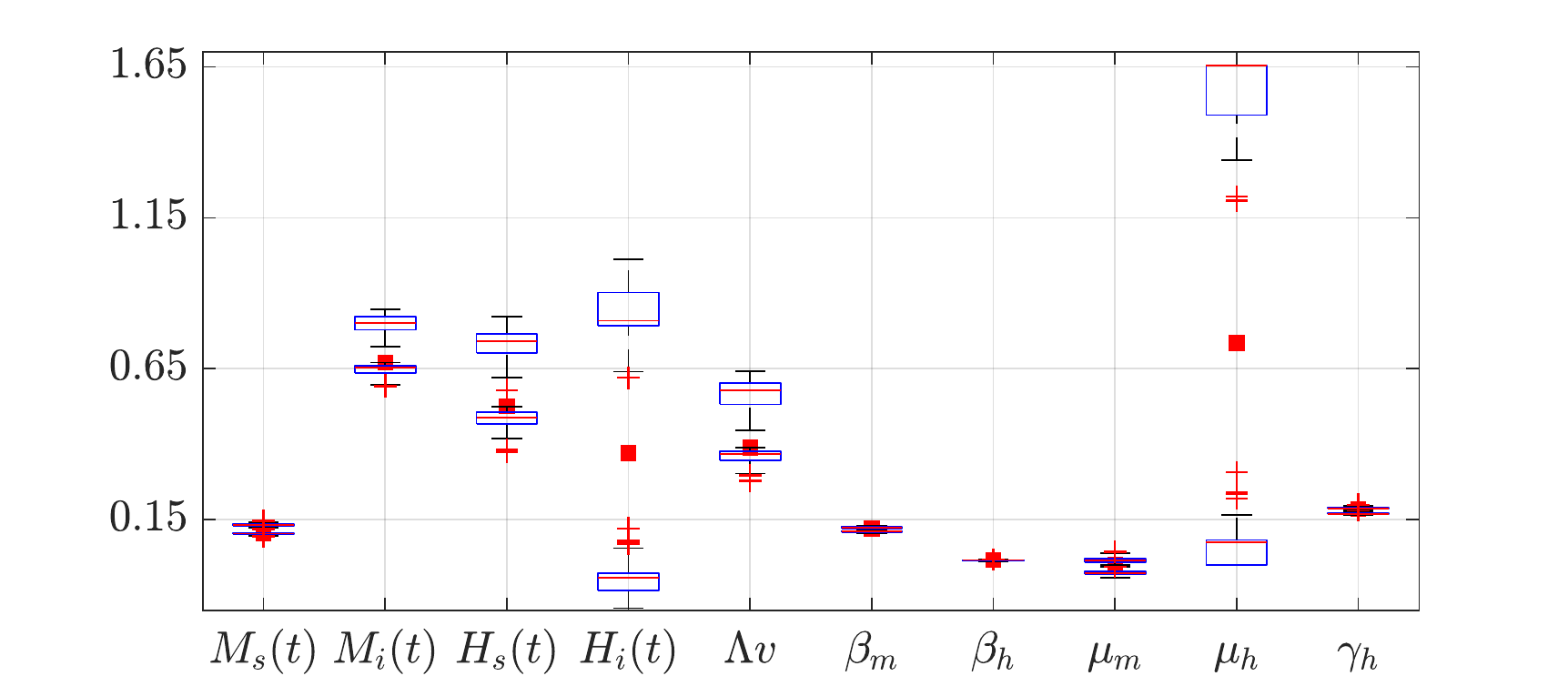}
        \end{subfigure}
        \caption{Boxplot for superior and inferior bounds of 100 CSB intervals. Each factor has in the same column the boxplot for both inferior and superior bounds, smaller boxplots indicate high accuracy for the interval, while large boxplots suggest stochastic shrinks in the factor interval estimation. Note that CSB requires a promissory search box from Algorithm \ref{al:oat}. This new search box could be greater than the original one, hence Algorithm \ref{al:UCI} may estimate intervals that overpasses those in the factor estimation task.} \label{fig:100csb}
    \end{figure}

To assess the uniqueness condition of the CSB we performed 100 CSB estimations processes, then, we normalize the got interval bounds for each factor regarding its estimation interval and present the inferior and superior bounds, from the 100 estimations, as a boxplot in Figure \ref{fig:100csb}. We found that there are stochastic shrinks, however, results are consistent for most of the factors.

\subsubsection{Sensitivity analysis for confidence sub-contour}
    
\begin{figure}[htbp!]
        \centering
        \begin{subfigure}[ht]{0.65\textwidth}
            \includegraphics[width=1\textwidth]{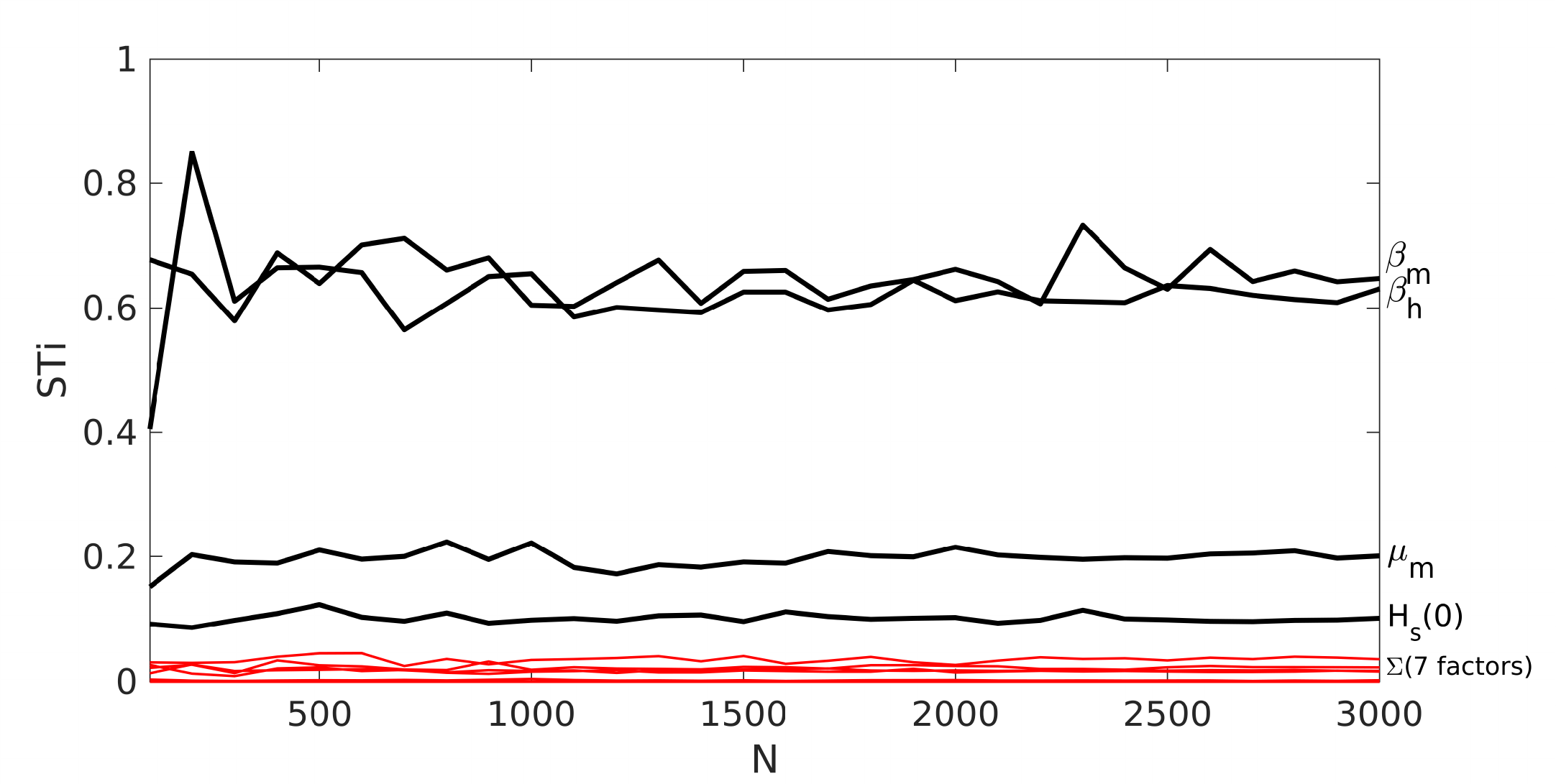}
            \centering
            \caption{Convergence analysis for total-order sensitivity indices ($ST_i$) of each factor in model \eqref{DengueModel}}
        \end{subfigure}
        \hfill
        \centering
        \begin{subfigure}[ht]{0.65\textwidth}
            \includegraphics[width=1\textwidth]{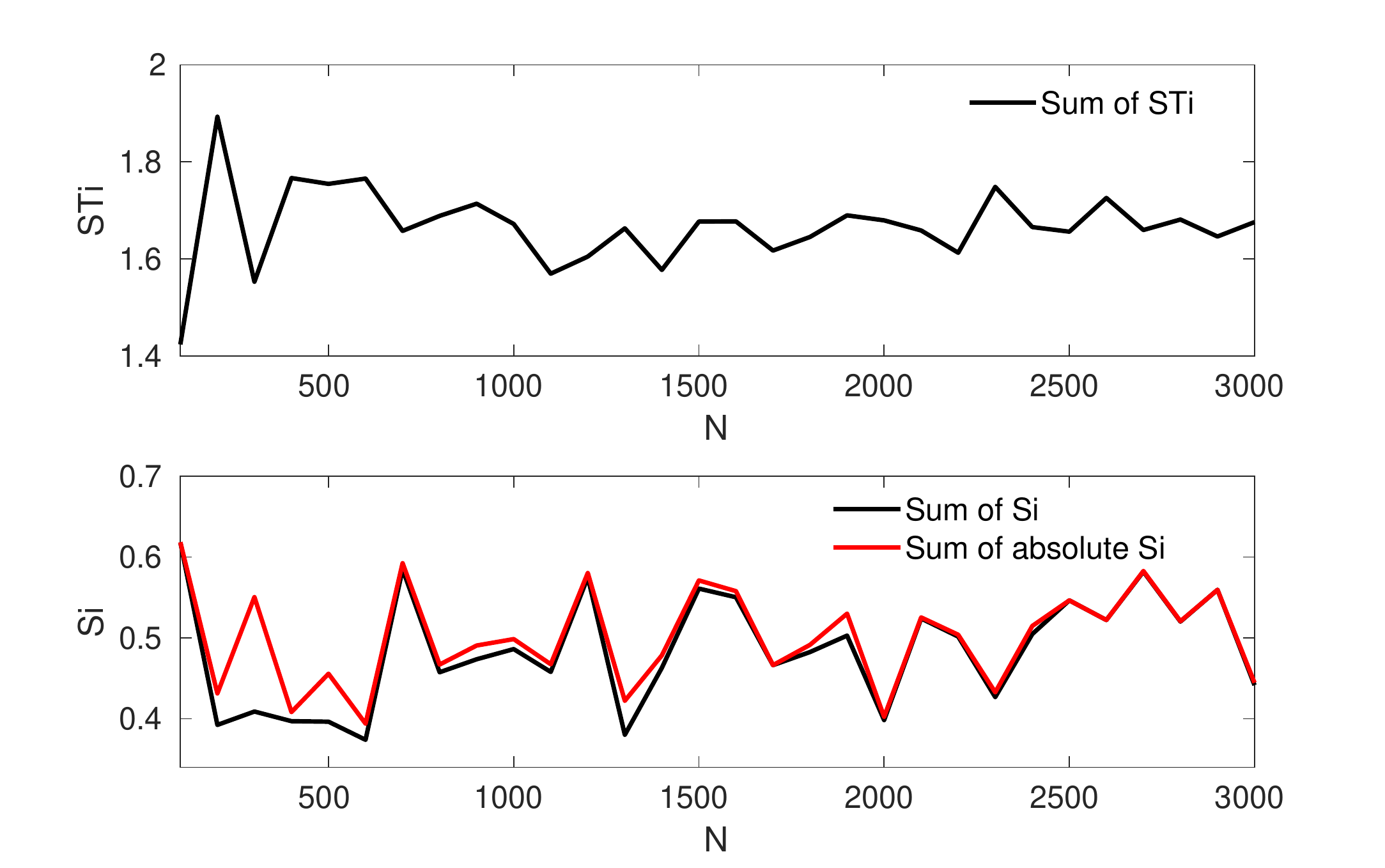}
            \caption{Convergence analysis for summation of total-order sensitivity indices ($ST_i$), first-order sensitivity indices ($S_i$) and absolute value of $S_i$}
        \end{subfigure}
        \caption{Three graphical criteria to select a sample size ($N$) that guarantee informative SA. Since it is unlikely to explore the whole space of factors through samples, we recommend as the best criterion the agreement between $S_i$ and absolute $S_i$, it is, a sample size such that the sensitivity estimators are reliable.}\label{fig:convergence}
    \end{figure}

    From convergence results in Figure \ref{fig:convergence} we chose a sample size ($N$) of 3000, which is supported since it meets all the implemented graphical criteria for convergence. Note that the summation of first-order sensitivity indices $S_i$ do not show a clear convergence to a specific value, however, for $N>2000$ there is a high coincidence between $S_i$ an absolute $S_i$ summation. From above, even a sample size of 2000 should be enough to get reliable estimations of the sensitivity indices, which is our real goal.

  \begin{figure}[htbp!]
        \centering
        \begin{subfigure}[ht]{0.4\textwidth}
           \includegraphics[width=1\textwidth]{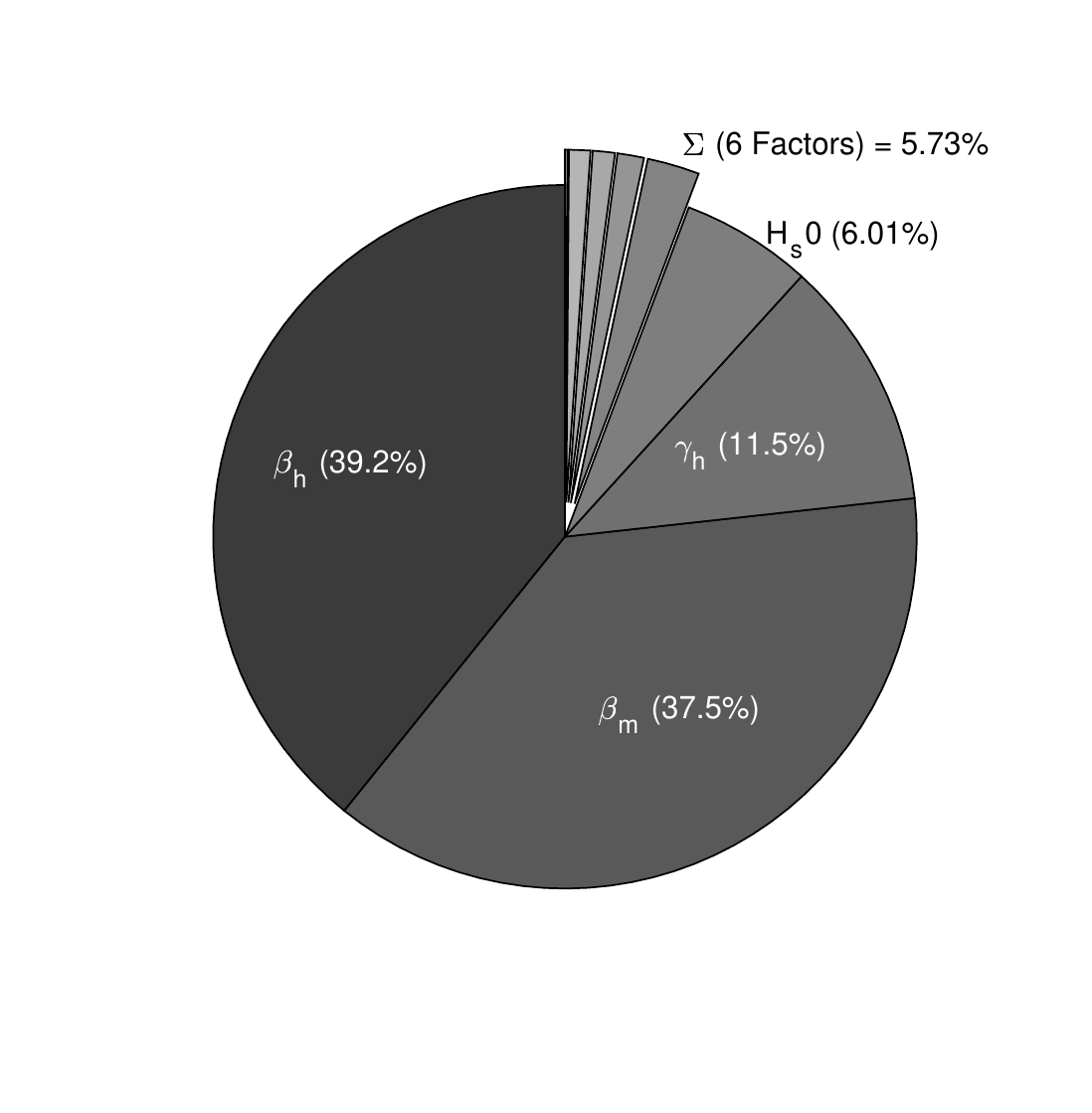}
            \caption{SA for estimation ranges (search box).}\label{fig:sa_er}
            \centering
            
        \end{subfigure}
        \centering
        \begin{subfigure}[ht]{0.4\textwidth}
            \includegraphics[width=1\textwidth]{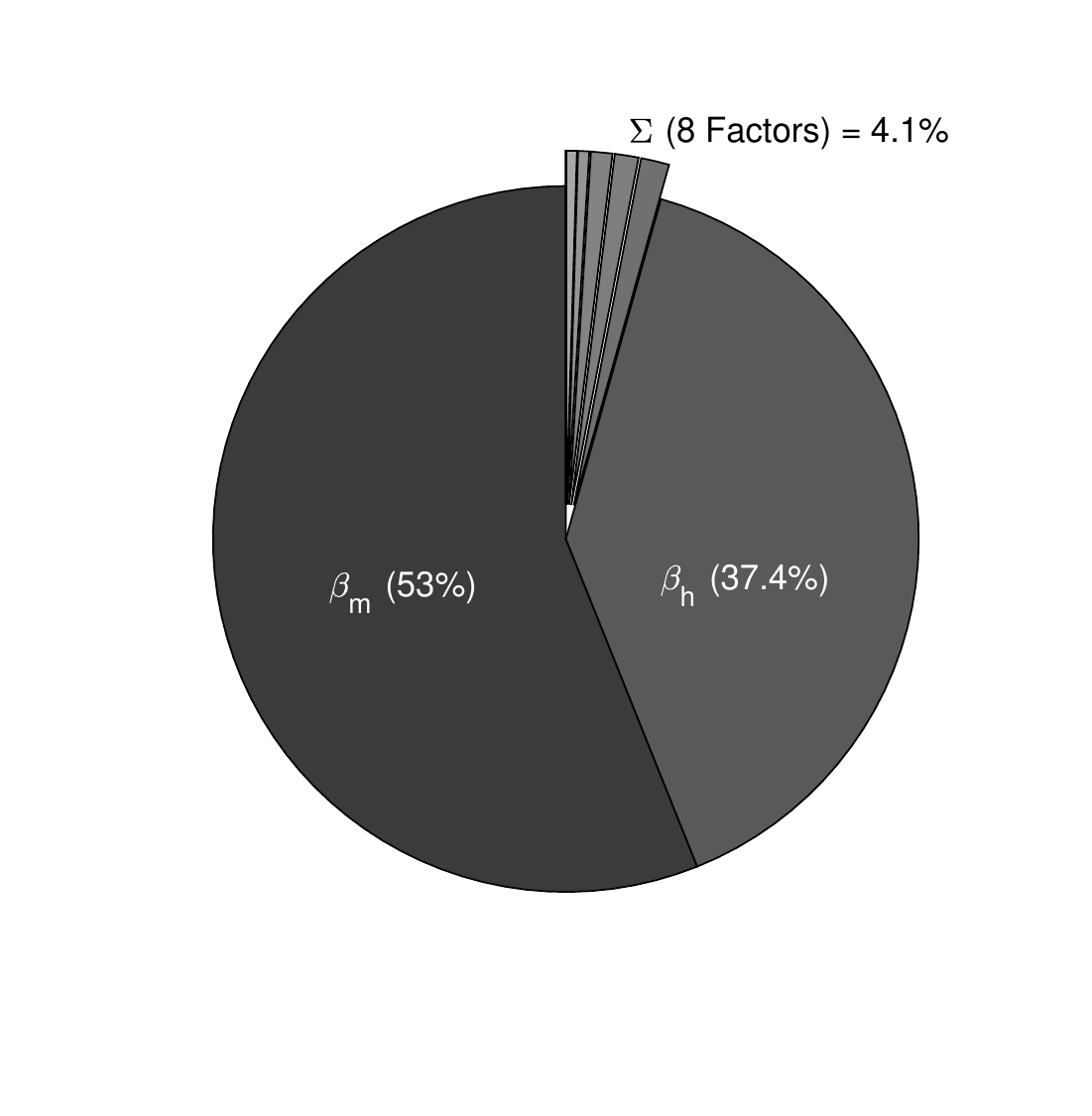}
            \caption{SA for traditional median CI.}\label{fig:sa_CI}
        \end{subfigure}
        \centering
        \begin{subfigure}[ht]{0.4\textwidth}
            \includegraphics[width=1\textwidth]{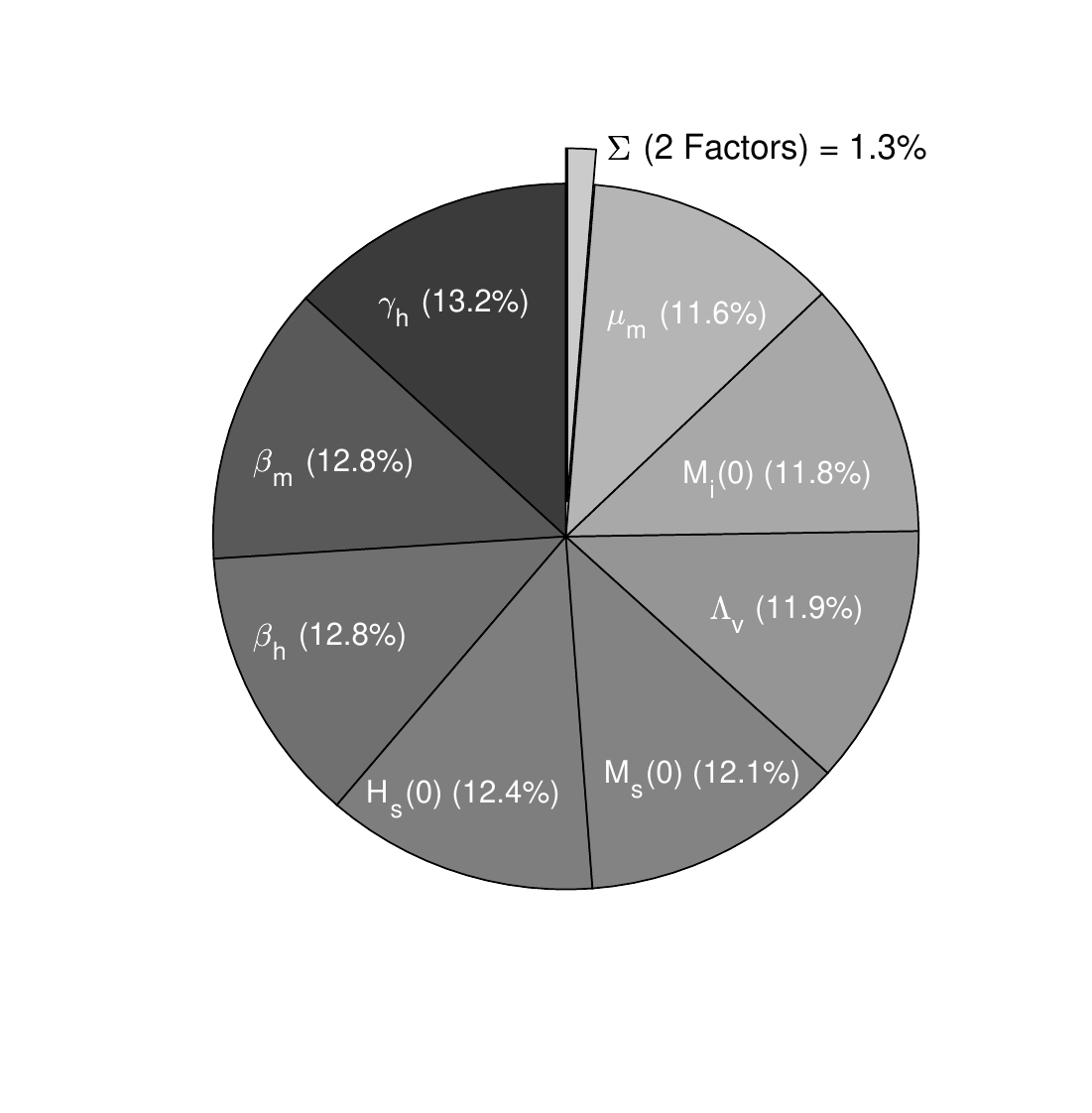}
            \caption{SA for the promissory search box.}\label{fig:sa_oat}
        \end{subfigure}
        \centering
        \centering
        \begin{subfigure}[ht]{0.4\textwidth}
            \includegraphics[width=1\textwidth]{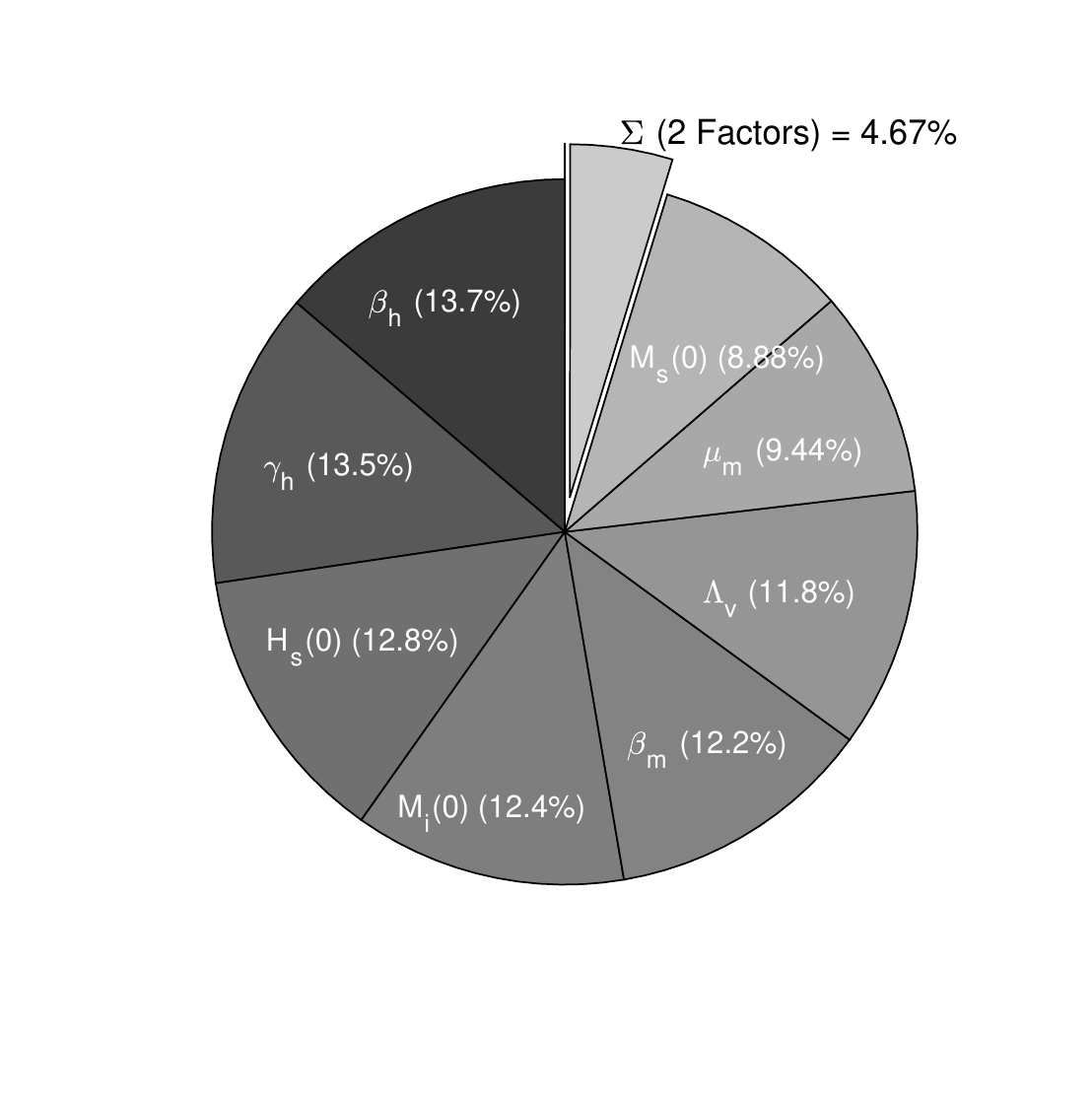}
            \caption{SA for CSB.}\label{fig:sa_csb}
        \end{subfigure}
        \caption{SA for different intervals that take part on the CSB estimation or in traditional median CI estimation. Intervals given to the \textit{gsua\_sa} method for sensitivity indices calculation are presented in Table \ref{Table:Results}, except for Figure \ref{fig:sa_oat}.}\label{fig:sa}
    \end{figure}
    
    Figure \ref{fig:sa} is the most relevant result from this section of application case. From Figure \ref{fig:sa_er} we know that dissimilarity contribution in the search box (estimation ranges, Table \ref{Table:Results}) is principally due to $\beta_m$ and $\beta_h$ factors, i.e., factor estimation mostly depends on the values for those factors. Thereby, we could expect $\beta_m$ and $\beta_h$ to be precisely estimated. From Figure \ref{fig:sa_CI} we see that even filtering and selecting the best estimation for factors, $\beta_m$ and $\beta_h$ still are the major contributors to dissimilarity, and moreover, they almost uniquely determine that dissimilarity. Those factors can be unidentifiable indeed, i.e., they have a correlation that allows model \eqref{DengueModel} getting a good fit to real data through multiple values for $\beta_m$ and $\beta_h$. Further, as we link the results from uncertainty analysis (Figure \ref{fig:ua}) with SA (Figure \ref{fig:sa_CI}) for traditional CI, we realize that: taking into account the number of factor estimations we selected and the characteristics of the median as a centrality measure, we should have got an uncertainty analysis with all model outputs closer to the nominal one, except we have identifiability issues in the model.
    
    From Figure \ref{fig:sa_oat}, we realize that there are no strong interactions among factors for the selected uncertainty level $\Lambda=30\%$. It is a little bit contradictory with the model identifiability suggestion from the above paragraph, however, different intervals for SA provides different results \cite{Saltelli2008}. Note that SA in Figure \ref{fig:sa_oat} suggest that two of the ten considered factors for model \eqref{DengueModel} have a minimum contribution to dissimilarity. The above does not suggest these factors to be irrelevant for the model. It suggests that those two factors do not contribute to dissimilarity as much as the other factors, in the context of the current search box.
    
    In Figure \ref{fig:sa_csb}, we present the results of the SA for CSB intervals in Table \ref{Table:Results}. It is noticeable that as we expected, the contribution of factors to dissimilarity is more uniform, i.e., it is an increment in the contribution of one of the two irrelevant factors in Figure \ref{fig:sa_oat}. However, there is still a factor that does not contribute to the dissimilarity in the CSB. At this point, it is necessary to mention a minor issue that is turning itself into relevant information: when applying Algorithm \ref{al:oat} to get the promissory search box; we noted that the algorithm was not able to find the exact required values for $\mu_h$ (the irrelevant factor), neither the superior nor the inferior bounds. At the start, we did not pay attention to this issue because it rises during the bisection search execution (Algorithm \ref{al:bisection}); which means that the factor can cause model outputs with dissimilarity beyond the threshold, but it is difficult to find a value that exactly meets the search criterion. As we analyze the information above and SA for CSB together, we can suggest that $\mu_h$ is a factor that does not determine the model output but it is still important to define both the model and the search box within the model has meaning.

\section{Final remarks}

We managed to propose a novel method for estimation of confidence intervals with acceptable computational cost and implement it in a Matlab toolbox. It is still necessary to prove the method with large models, synthetic conditions, and different parameterization; however, from the results presented in Section \ref{sec:appcase} we identified the following features:

\begin{itemize}
    \item The CSB successfully defines a $n-$orthotope within the confidence contour such that $\Delta\%$ of the model outputs are below a user-defined dissimilarity threshold.
    \item Since it is possible to some regions of the confidence contour being outside of the search box, it is likely to have some regions of the CSB outside the search box.
    \item The CSB intervals are strongly linked to sensitivity analysis theory; thus, large intervals correspond to less sensitive factors while small intervals suggest most relevant factors. To compare the length of the intervals, they should be normalized regarding the search box.
    \item Those factors that do not meet the same relevance in SA analysis for CSB must be linked to interactions beyond first-order indices. We suggest that since it is difficult to assess high order interactions from scatter plots.
    \item Close to zero total sensitivity indices should be interpreted as: those non-relevant factors could be still necessary to define the model, however, they can be fixed at any point within its search interval and it should not affect the process of factor estimation.
    \item The uncertainty level for CSB is user-defined, then, researchers can get confidence intervals that meet with uncertainty in real data.
    
\end{itemize}


\bibliography{mybibfile}

\appendix

\end{document}